\documentclass{ws-procs9x6}
\usepackage{makeidx}
\usepackage{latexsym} 
\usepackage{amssymb}  
\usepackage{epsfig}   

\newcommand{\bz}{{\bf z}}
\newcommand{\bp}{{\bf p}}

\newcommand{\beq}{\begin{equation}}
\newcommand{\eeq}{\end{equation}}
\newcommand{\bea}{\begin{eqnarray}}
\newcommand{\eea}{\end{eqnarray}}
\newcommand{\Tr}{\hbox{Tr}}
\newcommand{\C}{{\mathcal C}}       

\newcommand{\pa}{\parallel}
\newcommand{\pe}{\perp}
\newcommand{\dd}[1]{\mbox{d} #1}
\newcommand{\dddd}[1]{\dd{{}^{4} #1}}

\begin{document}

\title{Nonequilibrium quantum fields\\ with large fluctuations
}
\author{J{\"u}rgen Berges and Markus M.\ M{\"u}ller}
\address{Institute for Theoretical Physics, Heidelberg University,\\
        Philosophenweg 16, 69120 Heidelberg, Germany}
\maketitle
\abstracts{
We consider the nonequilibrium 
evolution of an \mbox{$O(N)$--symmetric} scalar quantum field theory 
using a systematic two--particle irreducible $1/N$--expansion to
next-to-leading order, which includes scattering and memory effects. 
The corresponding ``full Kadanoff-Baym equations'' are solved numerically 
without further approximations. This allows one to obtain a controlled
nonperturbative description of far-from-equilibrium dynamics and the 
late-time approach to quantum thermal equilibrium. Employing in addition a 
first-order gradient expansion for the Wigner transformed correlators 
we derive kinetic equations. In contrast to standard descriptions
based on loop expansions, our equations remain valid for
nonperturbatively large fluctuations. As an application, we discuss
the fluctuation dominated regime following parametric resonance 
in quantum field theory.}

In recent years we have witnessed an enormous increase of interest 
in the dynamics of quantum fields out of equilibrium. Strong motivation
in elementary particle physics comes, in particular, from 
current and upcoming relativistic heavy-ion collision experiments, 
phase transitions in the early universe or astrophysical applications.
Here the involved nonequi\-librium dynamics is often characterized by 
large corrections from quantum-statistical fluctuations that are
not accessible in a weak coupling or loop expansion. A paradigm
for such a situation is provided by the phenomenon of 
parametric resonance, which represents an important building block 
for our understanding of the (pre)heating of the early universe after a 
period of inflation.\cite{preheat1} In this context the resonant
amplification of fluctuations leads to explosive particle production
with a transition from a classical to a fluctuation dominated regime,
characterized by nonperturbatively large occupation numbers 
inversely proportional to the coupling.

Until recently\cite{BeSe}, classical field 
theory studies on the lattice have been the only quantitative 
approach available\cite{preheat2}. These are expected to be valid for 
not too late times, before the approach to quantum thermal equilibrium 
sets in. Calculations in quantum field theory had been 
limited to mean-field type approximations (leading-order 
in large-$N$, or Hartree)
which neglect scatterings.\cite{LOapp} These approximations are known 
to fail to describe late-time thermalization and, even at early times,  
do not give a valid description of the entire amplification regime
as pointed out in Refs.\ \cite{BeSe,preheat3}. 

In Ref.\ \cite{BeSe}
the first study of parametric resonance in quantum field theory 
from a next-to-leading order (NLO) calculation in a 
systematic two-particle irreducible (2PI) $1/N$--expansion 
\cite{Be1,AaAhBaBeSe} was presented, which includes scattering and memory 
effects.
The classical resonant amplification at early times 
is found to be followed by a collective amplification regime with
explosive particle production in a broad momentum range.
As a consequence, one observes rapid prethermalization
with a particle number distribution monotonous in momentum.
In particular, in this regime there are leading contributions
from all 2PI loop orders and standard weak coupling or loop expansions
break down. For its description it is crucial to employ 
a nonperturbative approximation as provided by the 
$1/N$--expansion at NLO.\cite{BeSe}

In this note, after reviewing the nonperturbative physics
involved in the phenomenon of parametric resonance,
we present suitable kinetic equations derived\cite{BeMu} from the
2PI $1/N$--expansion at NLO\cite{Be1}. For similar approximation
schemes see also Ref.\ \cite{IvKnVo3}.
Standard descriptions are typically based on a loop expansion of the 2PI
effective action. In particular the classical Boltzmann equation can be 
obtained starting from a three--loop approximation.\cite{BaKa,Da,CaHu}
The advantage of our kinetic equations is that their applicability 
is not limited by the question of the validity of a loop expansion.
For sufficiently large $N$, \cite{AaBe2}
it is only restricted by the applicability of a gradient expansion
employed in the derivation of the kinetic equations. Apart from
the example of the fluctuation dominated regime following
parametric resonance, as a further important application these equations 
are expected to be 
valid near second-order phase transitions. Here the correlation length 
diverges at the transition in the static limit. The possibility of 
enhanced fluctuations near 
such a critical point for the nonequilibrium dynamics has received much 
attention recently in the context of relativistic heavy-ion 
collisions.\cite{BeRa} A detailed discussion will be presented 
elsewhere.\cite{BeMu} \\ 

We consider a real scalar $N$--component quantum field 
$\varphi_a$ ($a\!=\!1,\ldots, N$) with $\lambda/(4! N)\, 
(\varphi_a\varphi_a)^2$
interaction, where summation over repeated indices is implied.
All correlation functions of the quantum theory can be obtained from the 
2PI generating functional for Green's 
functions $\Gamma[\phi,G]$, parametrized 
by the field expectation value 
$\phi_a(x)=\langle\varphi_a(x)\rangle$ 
and the connected propagator $G_{ab}(x,y)$: \cite{CoJaTo}
\bea
\Gamma[\phi,G] = S[\phi] + \frac{i}{2} \Tr\ln G^{-1} 
          + \frac{i}{2} \Tr\, G_0^{-1}(\phi)\, G
          + \Gamma_2[\phi,G] \, .
\nonumber
\eea 
Here $ i G^{-1}_{0,ab}(x,y;\phi) \equiv 
\delta^2 S[\phi]/\delta\phi_a(x)\delta\phi_b(y)$,
where $S$ is the classical action with 
$S_0 = - \int_x \phi_a(\square_x + m^2)\phi_a/2$ as the free part.  
We use the notation $\int_x \equiv \int_\C {\rm d}x^0 \int {\rm d}{\bf x}$
with $\C$ denoting a closed time path along the real axis.\cite{Ke}
\begin{figure}[t!]
  \centering
  \vspace*{-0.2cm}
  \epsfig{file=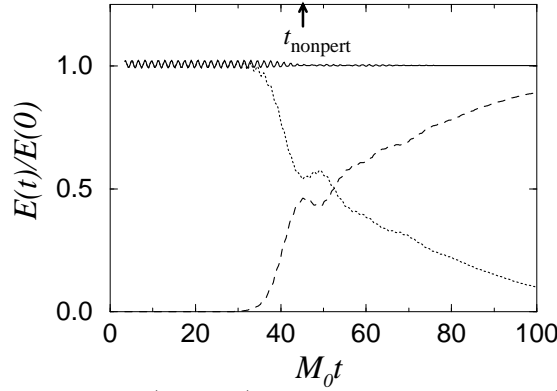,width=5.3cm,angle=-90}
  \vspace*{-0.4cm}
  \caption{\label{energy} Total energy $E_{\rm tot}$ (solid line) and 
  classical field energy $E_{\rm cl}$ (dotted line) as a function of time for
  $\lambda =10^{-6}$ and $N=4$ in $3+1$ dimensions, from Ref. ${}^2$.
  The dashed line represents the fluctuation part, 
  showing a transition between a classical field and a
  fluctuation dominated regime. For $t \gtrsim t_{\rm nonpert}$ 
  there are leading contributions from all loop orders and 
  the employed nonperturbative approximation becomes crucial.}
\end{figure}  
At NLO in the $1/N$--expansion of the 2PI effective action one 
has\cite{Be1,AaAhBaBeSe}
\bea 
\label{NLOcont} 
\Gamma_2[\phi,G] &=&  -\frac{\lambda}{4!N} \int_{x} G_{aa}(x,x)G_{bb}(x,x)
+ \frac{i}{2} \Tr \ln [\, {\bf B}(G)\, ]  \nonumber \\ 
&& + \frac{i\lambda}{6N} \int_{xy} 
{\bf I}(x,y;G) \phi_a(x) G_{ab}(x,y) \phi_b (y) \, ,\\
{\bf B}(x,y;G) &=& \delta_{\C}(x-y)
 + i \frac{\lambda}{6 N} G_{ab}(x,y)G_{ab}(x,y) \, , \label{BB}\\
{\bf I} (x,y;G) &=& \frac{\lambda}{6 N} G_{ab}(x,y) G_{ab}(x,y) \nonumber\\
 && - i \frac{\lambda}{6 N} \int_{z} {\bf I}(x,z;G)
 G_{ab}(z,y) G_{ab}(z,y) \, . \label{II}
\eea
Here the function ${\bf I}$ resums an infinite number of ``chain 
graphs''.\cite{Be1}
The equations of motion for $\phi_a$ and $G_{ab}$ are given by~\cite{CoJaTo}
\beq
\frac{\delta \Gamma[\phi,G]}{\delta \phi_a(x)} = 0 \quad, \quad
\frac{\delta \Gamma[\phi,G]}{\delta G_{ab}(x,y)} = 0 \, .
\label{fieldequations}
\eeq
The resulting evolution equations have been derived in detail
in Refs.~\cite{Be1,AaAhBaBeSe}. They have been solved numerically in 
Ref.\ \cite{BeSe} on a lattice along the lines of Ref.~\cite{Be1}.
We decompose the full two-point function using 
\beq
G_{ab} \left( x, y \right) = G_{ab}^> \left( x, y \right) \Theta_{\mathcal{C}} \left( x^0 - y^0 \right) + G_{ab}^< \left( x, y \right) \Theta_{\mathcal{C}} \left( y^0 - x^0 \right) 
\eeq
such that the spectral function $\rho$ and the symmetric propagator 
$F$ are given by \cite{Be1,AaBe}
\beq
\rho_{ab} \left( x, y \right) = - 2 {\rm Im} \Big( G_{ab}^> 
\left( x, y \right) \Big) 
\quad \mbox{and} \quad F_{ab} \left( x, y \right) = 
{\rm Re} \Big( G_{ab}^> \left( x, y \right) \Big) 
\eeq
\begin{figure}[t!]
  \centering
  \vspace*{-0.2cm}
  \epsfig{file=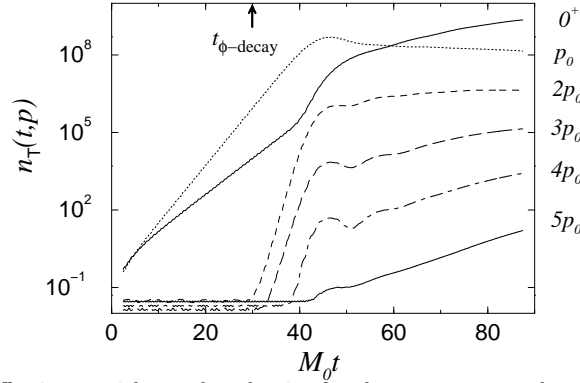,width=5.3cm,angle=-90}
  \vspace*{-0.4cm}
  \caption{\label{number_tr}Effective particle number density for 
  the transverse modes
  as a function of time for various momenta $0 \le p \le 5 p_0$ and
  same parameters as in Fig.\ \ref{energy}. At times 
  $t \lesssim t_{\rm nonpert}$ there is exponential amplification, which
  affects a broad momentum range for $t \gtrsim t_{\phi{\rm -decay}}$. 
  Taken from Ref. ${}^2$.}
\end{figure}  
For the discussion of parametric resonance in Ref.\ \cite{BeSe} a
system is considered, which is initially in a pure quantum state . Employing 
spatially homogeneous fields 
$\phi_a(t=t_0) \sim (6 N/\lambda)^{1/2}$,  
the initial propagator is taken to be diagonal with 
$F_{ab}= {\rm diag} \{F_{\pa},F_{\pe},\ldots,F_{\pe}\}$
and equivalently for $\rho_{ab}$. 
The effective transverse particle number density
\beq
\label{partnumb}
n_{\pe}(\bp) + \frac{1}{2} = \Big[ F_{\pe}(t,t';\bp) 
\partial_t\partial_{t'} F_{\pe}(t,t';\bp) \Big]^{1/2}_{t=t'} \, 
\eeq
and corresponding longitudinal particle number density $n_{\pa}$ are
defined as in Refs.\ \cite{Be1,AaBe}. Initially 
$n_{\pe}\equiv n_{\pa} \equiv 0$ such that the
initial conditions are characterized by zero 
particle number and large field amplitudes 
for small coupling $\lambda$. Correspondingly, the conserved total
energy $E_{\rm tot}$ is initially well approximated by the classical field 
contribution, i.e.\ $E_{\rm tot} \simeq E_{\rm cl}$, as shown in 
Fig.\ \ref{energy}. When the system evolves in time, 
more and more energy is converted into fluctuations. A detailed
discussion of the various characteristic regimes is given
in Ref.\ \cite{BeSe}.
For times $t \gtrsim t_{\rm nonpert}$ this leads
to nonperturbatively large particle number densities 
inversely proportional to the coupling. 
Neglecting the difference between transverse and longitudinal modes at
sufficiently late times, with $F_{ab} \simeq F\, \delta_{ab}$
and $\rho_{ab} \simeq \rho\, \delta_{ab}$ we have \cite{BeSe}
\beq
F \sim {\mathcal O}(N^0 \lambda^{-1})\,\, , \qquad
\rho \sim {\mathcal O}(N^0 \lambda^{0}) \, .
\label{estimate}
\eeq
The nonperturbatively large enhancement of the statistical 
propagator $F$ has the important consequence that any
approximation scheme based on a loop expansion of
the 2PI generating functional breaks down.  
We emphasize that the 2PI $1/N$--expansion at NLO remains valid as pointed 
out in \mbox{Ref.\ \cite{BeSe}}.
To discuss this in more detail, we consider here the simplified evolution 
equations for $F$ and $\rho$ with $\phi =0$. 
For the general case with a nonvanishing
field expectation value see Ref.\ \cite{BeSe}. One finds \cite{Be1,AaBe} 
\bea
\left[\square_x +M^2(x)\right]F(x,y)\! &=&
- \int_{t_0}^{x^0}\!\!\! {\rm d} z^0
\!\!\int \! {\rm d}\bz\,\, \Sigma_{\rho}(x,z)F(z,y) \nonumber \\
&&+ \int_{t_0}^{y^0}\!\!\! {\rm d} z^0 \!\!\int \!{\rm d}\bz\,\, \Sigma_{F}(x,z)
\rho(z,y), 
\label{eqF1}
\\ 
\left[\square_x +M^2(x)\right]\rho(x,y) &=&
-\int_{y^0}^{x^0}\!\!\! {\rm d} z^0 \!\!\int \! {\rm d}\bz\,\,
\Sigma_{\rho}(x,z)\rho(z,y).
\nonumber
\eea
At NLO in the $1/N$--expansion the effective mass term $M^2(x)$ is given by 
\beq
M^2(x) = m^2 +\lambda\frac{N+2}{6N}F(x,x)
\eeq
and the self--energies are \cite{Be1}
\bea
\Sigma_{F}(x,y) &=\!& -\frac{\lambda}{3N} \Big[ F(x,y)I_{F}(x,y)
- \frac{1}{4} \rho(x,y) I_\rho(x,y) \Big], \!\!\!
\label{sigmaF}\\
\Sigma_{\rho}(x,y) &=\!& -\frac{\lambda}{3N}
\Big[\rho(x,y)I_{F}(x,y)+F(x,y)I_{\rho}(x,y)\Big]. \!\!\!
\label{sigmarho}
\eea
Here the functions $I_{F}$ and $I_{\rho}$ contain the resummation: 
\bea
I_{F}(x,y) = 
-\frac{\lambda}{3}\Pi_{F}(x,y)
+\frac{\lambda}{3}
\int_{t_0}^{x^0}\!\!\! {\rm d} z^0
\int \! {\rm d}\bz\,\, I_{\rho}(x,z)\Pi_{F}(z,y)  
&&\nonumber \\
-\frac{\lambda}{3}
\int_{t_0}^{y^0}\!\!\! {\rm d} z^0 \int \!{\rm d}\bz\,\, I_{F}(x,z)\Pi_{\rho}(z,y),
&& \label{eqIF1}\\
I_{\rho}(x,y) = 
-\frac{\lambda}{3}\Pi_{\rho}(x,y)
+\frac{\lambda}{3}
\int_{y^0}^{x^0}\!\!\! {\rm d} z^0 \int \! {\rm d}\bz\,\,
I_{\rho}(x,z)\Pi_{\rho}(z,y),
&& \label{eqIrho1}
\eea
with
\bea
\Pi_{F}(x,y) &=& -\frac{1}{2}\Big(F^2(x,y)-\frac{1}{4}\rho^2(x,y)\Big),
\label{PIF}\\
\Pi_{\rho}(x,y) &=& -F(x,y)\rho(x,y).
\label{PIR}
\eea
Using (\ref{estimate}) in (\ref{eqIF1}) and (\ref{eqIrho1}), one observes that 
each term of the infinite resummation of loops contained in $I_F$ and $I_\rho$
contributes with the same order in $\lambda$. To describe this regime and 
the late-time behavior a nonperturbative approximation has to be employed, 
such as the 2PI $1/N$--expansion at NLO.\cite{Be1,AaAhBaBeSe}

In the fluctuation dominated regime, for $t \gtrsim t_{\rm nonpert}$, the 
dynamics is characterized by a slow drifting of modes\cite{Be1}, which 
suggests (cf. Ref. \cite{AaBe}) the applicability
of a first order gradient expansion with respect to the center coordinate 
$X = (x+y)/2$. \cite{IvKnVo3,BaKa,Da,CaHu,BlIa} Exploiting the effective 
loss\cite{Be1} of the dependence on 
the initial time for sufficiently large $x^0$ and $y^0$, we send 
$t_0 \to -\infty$ in Eqs. (\ref{eqF1}) and (\ref{eqIF1}). Using 
the retarded and the advanced propagator 
\vfill
\[ G_R(x, y) = \Theta \left( x^0 - y^0 \right) \rho (x, y) \;, \]
\vfill
\[ G_A(x, y) = - \Theta \left( y^0 - x^0 \right) \rho (x, y) \]
\vfill
as well as corresponding definitions for the retarded and advanced 
self--energies $\Sigma_{R,A}$ and the resummation functions $I_{R,A}$, one 
can then also send the upper limits to infinity and Fourier transform with 
respect to the relative coordinate $s=x-y$. For the two--point functions we
write
\vfill
\[ \tilde{F} \left( X, k \right) = \int \dddd{s} \; e^{i k s} F \left( X + \frac{s}{2}, X - \frac{s}{2} \right) \;, \]
\vfill
\[ \tilde{\varrho} \left( X, k \right) = - i \int \dddd{s} \; e^{i k s} \rho \left( X + \frac{s}{2}, X - \frac{s}{2} \right) \;. \]
\vfill
Here we introduced a factor $-i$ in the definition of the Wigner transformed 
spectral function to obtain a real $\tilde{\varrho} \left( X, k \right)$.
The advanced and the retarded propagators satisfy 
${\tilde{G}_R}^*(X,k) = \tilde{G}_A(X,k)$. Again, this property also holds for 
the corresponding self--energies. With the notation
\vfill
\[ \Big( \tilde{f} \ast \tilde{g} \Big) (X,k) \equiv \int \frac{{\rm d}^4 q}{(2\pi)^4}\, \tilde{f}(X,k-q)\, \tilde{g}(X,q) \]
\vfill
and the definition of the Poisson brackets
\vfill
\begin{eqnarray*}
      \left\{ \tilde{f} ( X, k) ; \tilde{g} ( X, k ) \right\}_{PB}
  &=& \Big[ \partial_{k_{\mu}}\tilde{f} ( X, k) \Big] \Big[\partial_{X^{\mu}} \tilde{g} ( X, k )\Big] \nonumber \\
  & & - \Big[\partial_{X^{\mu}} \tilde{f} ( X, k)\Big] \Big[\partial_{k_{\mu}}\tilde{g} ( X, k )\Big] \;,
\end{eqnarray*}
\vfill
\noindent one finds from a first--order gradient expansion of Eqs. (\ref{eqF1}) -- 
(\ref{PIR}) the kinetic equations:
\clearpage
\bea \lefteqn{
\Big[ 2 k^{\mu} \partial_{X^{\mu}} + 
\left( \partial_{X^{\mu}} M^2 \left( X \right) \right) 
\partial_{k_{\mu}} \Big]\, \tilde{F} \left( X, k \right) } \nonumber \\
& = & \tilde{F} \left( X, k \right) \tilde{\Sigma}_{\varrho} 
\left( X, k \right) - \tilde{\Sigma}_F \left( X, k \right) 
\tilde{\varrho} \left( X, k \right) \label{eqF2} \\ 
  & +  &  \left\{ \tilde{\Sigma}_F \left( X, k \right) 
; {\rm Re\,} \tilde{G}_R \left( X, k \right)  \right\}_{PB} 
+ \left\{ {\rm Re\,} \tilde{\Sigma}_R \left( X, k \right)  
; \tilde{F} \left( X, k \right) \right\}_{PB} \;,
\nonumber
\eea
\vfill
\bea \lefteqn{
\Big[ 2 k^{\mu} \partial_{X^{\mu}} + \left( \partial_{X^{\mu}} M^2 
\left( X \right) \right) \partial_{k_{\mu}} \Big]\, \tilde{\varrho} 
\left( X, k \right)} \\
& = & \left\{ \tilde{\Sigma}_{\varrho} \left( X, k \right) ; 
{\rm Re\,} \tilde{G}_R \left( X, k \right) \right\}_{PB} + 
\left\{ {\rm Re\,} \tilde{\Sigma}_R \left( X, k \right) ; 
\tilde{\varrho} \left( X, k \right) \right\}_{PB} \;.
\nonumber
\eea
Here the Wigner transformed self--energies are given by
\begin{equation}
  \tilde{\Sigma}_F \left( X, k \right) = - \frac{\lambda}{3 N} \left( \left( \tilde{F} \ast \tilde{I}_F \right) \left( X, k \right) + \frac{1}{4} \left( \tilde{\varrho} \ast \tilde{I}_{\varrho} \right) \left( X, k \right) \right) \;,
\end{equation}
\vfill
\begin{equation}
  \tilde{\Sigma}_{\varrho} \left( X, k \right) = - \frac{\lambda}{3 N} \bigg( \left( \tilde{F} \ast \tilde{I}_{\varrho} \right) \left( X, k \right) + \left( \tilde{\varrho} \ast \tilde{I}_F \right) \left( X, k \right) \bigg)
\end{equation}
and the equations for the resummation functions read
\bea \lefteqn{
 \tilde{I}_{\varrho} \left( X, k \right)
   =   - \frac{\lambda}{3} 
\Bigg( \tilde{\Pi}_{\varrho} \left( X, k \right)  
\left[ 1 - \tilde{I}_R \left( X, k \right) \right] 
  + \tilde{I}_{\varrho} \left( X, k \right) 
\left( \tilde{G}_A \ast \tilde{F} \right) \left( X, k \right) }
\nonumber \\
  &   &  - \frac{i}{2} \left\{ \tilde{I}_R 
\left( X, k \right) ; \tilde{\Pi}_{\varrho} 
\left( X, k \right) \right\}_{PB} 
 + \frac{i}{2} \left\{ \tilde{I}_{\varrho} 
\left( X, k \right) ; \left( \tilde{G}_A \ast \tilde{F} \right) 
\left( X, k \right) \right\}_{PB} \Bigg) \;,
\nonumber \\
\eea
\vfill
\bea \lefteqn{
  \tilde{I}_F \left( X, k \right)
  =  - \frac{\lambda}{3} \Bigg( \tilde{\Pi}_F \left( X, k \right) 
\left[ 1 - \tilde{I}_R \left( X, k \right) \right] 
  + \tilde{I}_F \left( X, k \right) \left( \tilde{G}_A \ast 
\tilde{F} \right) \left( X, k \right) } 
\nonumber \\
  &   &  + i \left\{ \tilde{I}_R \left( X, k \right) ; 
\tilde{\Pi}_F \left( X, k \right) \right\}_{PB} 
         + \frac{i}{2} \left\{ \tilde{I}_F 
\left( X, k \right) ; \left( \tilde{G}_A \ast \tilde{F} \right) 
\left( X, k \right) \right\}_{PB} \Bigg) \;.
\nonumber\\
\eea
Here $\tilde{\Pi}_{F, \varrho} \left( X, k \right)$ are the Wigner 
transforms of the functions defined in Eqs.~(\ref{PIF}) and (\ref{PIR}).
At this order of the gradient expansion the retarded propagator fulfills
the algebraic equation
\begin{equation}
  \tilde{G}_R \left( X, k \right) = \Big[ - k^2 + M^2 \left( X \right) + \tilde{\Sigma}_R \left( X, k \right) \Big]^{-1} \;.
\end{equation}
Similarly the retarded self--energy and the retarded resummation function 
satisfy
\clearpage
\beq
\tilde{\Sigma}_R \left( X, k \right) = - \frac{\lambda}{3 N} 
\Big[ \left( \tilde{F} \ast \tilde{I}_R \right) 
\left( X, k \right) + \left( \tilde{G}_R \ast \tilde{I}_F \right) 
\left( X, k \right) \Big]
\eeq
and
\bea
 \tilde{I}_R \left( X, k \right) 
  & = & \frac{\lambda}{3} \left[ 1 - \tilde{I}_R 
\left( X, k \right) \right] \left( \tilde{G}_R \ast \tilde{F} \right) 
\left( X, k \right) \nonumber \\
  &   &  - \frac{i \lambda}{6} \left\{ \tilde{I}_R \left( X, k \right), 
\left( \tilde{G}_R \ast \tilde{F} \right) \left( X, k \right) \right\}_{PB} \;. \label{eqIR2}
\eea
We emphasize that Eqs.~(\ref{eqF2}) -- (\ref{eqIR2}) represent a closed set of
equations. In their range of applicability these equations have the advantage
that they do not involve an integration over time history. This is
important for an efficient description of the late--time behaviour of quantum
fields. A detailed discussion will be presented in 
Ref.\ \cite{BeMu}. \\[0.5ex]

We thank Julien Serreau for collaboration on related work
\cite{BeSe,AaAhBaBeSe} and many discussions.

\printindex
\end{document}